\documentclass[aps,pre,twocolumn]{revtex4}
\usepackage{amsfonts,amssymb,amsmath,bm}
\usepackage{graphicx}
\usepackage{color}

\usepackage{tikz}
\usetikzlibrary{decorations.pathmorphing}
\usetikzlibrary{patterns,arrows}
\usetikzlibrary{decorations.markings}
\usepackage{pgfplots}
\usepackage{tkz-fct}

\newcommand{\I}{{i}}

\newcommand{\e}{\eqref}

\begin{document}

\title{Statistical properties of the laser beam propagating in a turbulent medium}

\author{Igor Kolokolov$^{1,2}$, Vladimir Lebedev$^{1,2}$, and Pavel M. Lushnikov$^{1,3}$}

\affiliation{$^1$ Landau Institute for Theoretical Physics, RAS, \\
142432, Chernogolovka, Moscow region, Russia \\
$^2$ NRU Higher School of Economics, \\
101000, Myasnitskaya 20, Moscow, Russia \\
$^3$  University of New Mexico, Albuquerque, NM 87131, USA }

\begin{abstract}

We examine statistical properties of a laser beam propagating in a turbulent medium. We prove that the intensity fluctuations at large propagation distances  possess Gaussian probability density function and establish quantitative criteria  for realizing the Gaussian statistics depending on the laser propagation distance, the laser beam waist, the laser frequency and the turbulence strength. We calculate explicitly the laser envelope pair correlation function and corrections to its higher order correlation functions breaking Gaussianity. We discuss also statistical properties of the brightest spots in the speckle pattern.

\end{abstract}


\maketitle

\section{Introduction}

Propagation of either the acoustic beam or the electromagnetic beam in a turbulent medium  is the classical problem which has been a subject of numerous investigations which especially intensified after the advent of lasers in 1960-ies. The shape of the  beam wavefront is progressively disturbed with the increase of the propagation distance due to the turbulent fluctuations. These distortions are random due to the chaotic nature of turbulence. Therefore they should be described statistically. Below we refer specifically to the laser beam while assuming that the same theory can be applied for the acoustic beam.

The early advances in beam propagation through turbulent medium outlined in Refs.
\cite{TatarskiiBook1961,TatarskiiBook1971,Goodman,StrohbehnBook1978} were focused on lower order statistical moments of the laser intensity. Among such moments often a scintillation index  $\sigma_I^2\equiv \langle I^2\rangle/\langle I\rangle^2-1$ is used. The index $\sigma_I^2$ is the measure of the strength of the fluctuations of the irradiance $I$ (the laser beam intensity) at the target plane. Here and below we denote by $\langle \ldots \rangle$ an average over the ensemble of the atmospheric turbulence realizations or, equivalently, over time.

At relatively small propagation distances, where irradiance fluctuation are small ($\sigma_I^2\ll1$), the classic perturbative approach well describes modification of the laser beam propagation due to turbulence \cite{TatarskiiBook1961,TatarskiiBook1971}. Statistically averaged beam characteristics
at larger distances in strong  scintillation regimes ($\sigma_I^2\gg 1 $) were addressed through semi-heuristic theory \cite{AndrewsPhillipsBook1998}. At such distances, the laser beam disintegrates into speckles with lower order statistical moments providing only very limited information about structure of the intensity fluctuations. It was found in Ref. \cite{LachinovaVorontsovJOpt2016} that
a significant fraction of deviation between the theoretical value of $\sigma_I^2$ \cite{AndrewsPhillipsBook1998} and simulations is due to rare large fluctuations of laser beam intensity. Such giant fluctuations were also observed in numerical experiments
\cite{VorontsovEtAlAMOSConf2010,VorontsovEtAlAMOSConf2011}. The work \cite{LushnikovVladimirovaJETP2018} studied the structure of large fluctuations and proposed to use them for the efficient delivery of laser energy over long distances by triggering the pulse laser operations only during the times of such rare fluctuations. It was demonstrated in  Ref.
\cite{LushnikovVladimirovaJETP2018} that after $7$ km propagation of the laser beam with the initial beam waist $1.5$ cm in typical atmospheric conditions, about $0.1\%$ atmospheric realizations carry $\gtrsim 28\%$ of initial power in a single giant fluctuation (a single speckle) of the laser intensity.

In this paper we address the statistical properties  of the laser intensity fluctuations and identify analytically the spatial profiles of the intense laser speckles. We consider a propagation of the initially Gaussian beam at large distances  corresponding to $\sigma_I^2\gg 1 $. We assume that the typical transverse size of speckles is much smaller than the Gaussian beam width which is also ensured at large enough distances of propagation through turbulent medium. We use the ladder sequence of diagrams of the stochastic perturbation theory  to  show that   the probability density function (PDF) of laser intensity fluctuations at such distances  is well approximated by the Gaussian stochastic process. The  properties of such stochastic process at each propagation distance $z$  is fully determined by the pair correlation function of the laser intensity fluctuation which has the explicit expression in the integral form. The spatial structure of large fluctuations (the bright spots of the laser intensity) of such stochastic process has the form of the transverse correlation function of the Gaussian process with the transverse width scaling as $\propto z^{-3/5}$ for the Kolmogorov-Obukhov spectrum of the atmospheric turbulence \cite{TatarskiiBook1961,TatarskiiBook1971}.  We also found non-Gaussian corrections to the higher order correlation functions and established that these corrections decrease with the propagation distance.

The plan of the paper is the following. In Section \ref{sec:basiceq} we
introduce basic equations in the physical variables, define the
fluctuations of the refractive index (Section  \ref{sec:Fluctuationsn1})
and discuss the characteristic propagation lengths which controls the
properties of laser beam fluctuations (Section \ref{sec:Spatialscales}).
Section \ref{sec:dimensionless} introduces   scaled dimensionless
variables. Section \ref{sec:Gaussianfluctuations} provides a reduction of
the refractive index fluctuations to the Gaussian stochastic process with the
power law pair correlation function. Section  \ref{sec:Parametersregions}
discusses the parameters and different spatial regions in the dimensionless
variables. Section \ref{sec:corr} analyzes the pair, fourth and higher order
 correlation functions as well as address the statistic of the
intensity fluctuations. Section \ref{sec:Effective action} develops the ladder
approximation as well as discusses the corrections beyond the ladder
approximation identifying the region of the applicability of the Poisson statistics.
Section \ref{sec:Conclusion} summarize the results and discuss their
applicability conditions. Appendices provide the details of the ladder
approximation, corrections beyond it and the calculations of the
statistics of the large fluctuations (bright spots) of the laser intensity.

\section{Basic equations}
\label{sec:basiceq}

A  propagation of a monochromatic  beam with a single polarization through  turbulent media  is described by the linear Schr\"odinger equation (LSE) (also called as Leontovich equation)
\cite{TatarskiiBook1961,TatarskiiBook1971,VlasovPetrishchevTalanovRdiofiz1971} for the spatiotemporal envelope of the electric field
 \begin{eqnarray}\label{nlsdimensionall}
  \I \frac{\partial }{\partial Z}\Psi+\frac{1}{2k_{0}}\nabla_\perp^2\Psi
  +k _{0}n_1\Psi=0,
 \end{eqnarray}
where all quantities are functions of ${\bm R}_\perp,Z,t$. Here the beam is aligned along $Z$-axis, ${\bm R}_\perp\equiv (X,Y)$ are the transverse coordinates, $t$ is the time, $\nabla_\perp\equiv(\partial _X, {\partial _Y}),$  $k_0=2\pi n_0/\lambda_0$ is the wavevector in the medium, $\lambda_0$ is the wavelength in the vacuum, $n=n_0+n_1$ is the linear index of refraction with the average value $n_0$ and the fluctuation contribution $n_1$, with zero average $\langle n_1 \rangle=0$. The beam intensity is expressed as $I=|\Psi|^2$.

We assume that the inverse time of the laser beam propagation is much larger than characteristic rates of the turbulent motions. Then corrections related to the dependence of $n_1$ on $t$ are negligible and we can operate in terms of flash realizations of the refractive index $n_1$. Statistics of such flash realizations can be characterized by simultaneous correlation functions of $n_1$. Thus $t$ does not explicitly enter Eq. \e{nlsdimensionall}, but serves as the parameter  distinguishing different atmospheric realizations, so below we omit $t$ in arguments of all functions. The neglect of the temporal derivative (that derivative results from the chromatic dispersion) is justified for laser beams with typical durations longer than a few nanoseconds.

Linear absorbtion (results in exponential decay of the laser intensity with propagation distance) is straightforward to include into Eq. \e{nlsdimensionall}. Kerr nonlinearity can be also added to Eq.
\e{nlsdimensionall} resulting in nonlinear Schr\"odinger  Eq. which describes the catastrophic self-focusing (collapse) of laser beam for laser powers $P$ above critical power $P_c$ ($P_c\sim3$GW for $\lambda_0= 1064\text{nm)}$
\cite{VlasovPetrishchevTalanovRdiofiz1971,ZakharovJETP1972,LushnikovDyachenkoVladimirovaNLSloglogPRA2013}
and multiple filamentation for $P\gg P_c$ \cite{LushnikovVladimirovaOptLett2010}. At distances well below the nonlinear length, one can consider Kerr nonlinearity as perturbation (see, e.g., Ref. \cite{VasevaFedorukRubenchikTuritsyn}) combining it with the effect of atmospheric turbulence. Such additions are beyond the scope of this paper.

We consider propagation of the beam produced by a laser located at $Z=0$. We assume that at the laser output the beam has a Gaussian shape
\begin{equation}\label{Gaussianbeam}
\Psi_{in}=A\exp\left(- R_\perp^2/w_0^2\right),
\end{equation}
where $w_0$ is the initial Gaussian beam waist, $A$ is the initial beam amplitude. The expression (\ref{Gaussianbeam}) should be treated as the initial condition to the equation (\ref{nlsdimensionall}) posed at $Z=0$.

At small $Z$ one can neglect the term with $n_1$ in Eq. (\ref{nlsdimensionall}). Then we find the explicit solution of Eq. (\ref{nlsdimensionall}) with the initial condition (\ref{Gaussianbeam})
given by the diffraction-limited Gaussian beam  \begin{equation}
 \Psi=\frac{w_0^2}{w_0^2+2iZ/k_0}
 \exp\left(-\frac{R_\perp^2}{w_0^2+2iZ/k_0}\right).
 \label{freesolution}
 \end{equation}
Note that at distances $Z\gg Z_R$, where $Z_R$ is the Rayleigh length
 \begin{equation}
 Z_R=k _0w_0^2/2,
 \label{Rayleighlength}
 \end{equation}
the beam width can be estimated as $Z/(k_0w_0)\gg w_0$, whereas the phase varies on a much smaller transverse spatial scale $\sim \sqrt{Z/k_0}$.

\subsection{Fluctuations of the refractive index}
\label{sec:Fluctuationsn1}

The refractive index variation $n_{1}$ is proportional to the density fluctuation of the turbulent medium. Description of principal properties of the atmospheric turbulence can be found in monograph \cite{Monin}. Typically, $l_0$ is in the range of few mm or even smaller, while $L_0$ is  ranging from many meters to kilometers. The main contribution to $n_{1}$ stems from the turbulent fluctuations with scales of the order of the integral length (the outer scale) of the turbulence $L_0$. We are however interested in atmospheric fluctuations on scales of the order of the laser beam width, that is assumed to be much smaller than $L_{0}$.
In terms of Eq. \e{nlsdimensionall} such fluctuations on the scale $L_0$\ result in the  change of the phase of $\Psi$ which are nearly homogenous and do not affect the laser intensity. Besides, the beam width is assumed to be much larger than Kolmogorov scale (the inner scale of turbulence) $l_0$. Then $l_0$ does not significantly affect PDF of laser beam fluctuations. In this situation the turbulent fluctuations relevant for the problem belong to the inertial scale of turbulence where they possess definite scaling properties \cite{Frisch} and are both homogeneous and isotropic.

Atmospheric fluctuations can be characterized by the structure function of the refractive index fluctuations that is the simultaneous average $\langle [n_1(\bm R_\perp,Z)-n_1({\bm 0},0)]^2\rangle$.  Remind that the angular brackets $\langle \ldots \rangle$ mean averaging over realizations of $n_1$ or, equivalently, averaging over time. In the inertial range of turbulence the Kolmogorov-Obukhov law is valid
 \begin{equation} \label{KolmogorovObukhov}
 \langle [n_1(\bm R_\perp,Z)-n_1({\bm 0},0)]^2\rangle
 =C_n^2 \rho^{2/3},
 \end{equation}
where $\rho=\sqrt{R_\perp^2+Z^2}$ and the factor $C_n^2$ characterizes strength of the turbulence. The expression (\ref{KolmogorovObukhov}) is correct provided $l_0\ll \rho\ll L_0$.

We now consider a little bit more general case not restricting to the Kolmogorov-Obukhov law \e{KolmogorovObukhov}. It is still natural to assume homogeneity and isotropy of the turbulence on scales smaller than its integral scale $L_0$. Then the structure function $\langle [n_1(\bm R_\perp,Z)-n_1({\bm 0},0)]^2\rangle$ depends solely on $\rho$, which is the absolute value of the separation between  points. We assume the following general power law coordinate dependence on scales from the inertial interval:
\begin{equation} \label{KolmogorovObukhovgeneral}
 \langle [n_1(\bm R_\perp,Z)-n_1({\bm 0},0)]^2\rangle
 =C_n^2 \rho^{\mu},
\end{equation}
where $\mu<1$ is the scaling exponent, characterizing the refractive index fluctuations. The particular case of the Kolmogorov-Obukhov law \e{KolmogorovObukhov} corresponds to $\mu=2/3$.

Using Eq. \e{KolmogorovObukhovgeneral}, one can represent the pair correlation function of $n_1$ by the expansion in three dimensional Fourier harmonics as follows (see e.g. Refs. \cite{TatarskiiBook1961,TatarskiiBook1971})
 \begin{align}
&  \langle n_1(\bm R_\perp,Z)n_1({\bm 0},0)\rangle
=\frac{{  C_n^2 \Gamma(\mu+2)}}{4\pi^2}\sin{\left
(\frac{\pi\mu}{2} \right )}
 \nonumber\\
& \times\int {d^2 q\,dq_z}
 \exp(i\bm q \cdot\bm R_\perp+i q_z Z)
 (q^2+q_z^2)^{-3/2-\mu/2},
 \label{paircorr2dimensional}
 \end{align}
where $q^2=q_x^2+q_y^2$. Here we assumed that the divergence of the integral at small $q,q_z$ is removed by a cutoff at the large scale $\rho\sim L_0$ which can be also taken care by considering the structure function \e{KolmogorovObukhovgeneral} instead of the pair correlation function.

The propagation length $Z$ is assumed to be much larger than the beam width, i.e. one can use the paraxial approximation with the characteristic wavevector $q$ much larger than the characteristic component $q_z$. Therefore $q_z$ can be neglected in Eq. (\ref{paircorr2dimensional}) in comparison with $q$. Then one obtains by replacing $q^2+q_z^2\to q^2$ that
 \begin{align}
&  \langle n_1(\bm R_\perp,Z)n_1({\bm 0},0)\rangle
=\frac{{  C_n^2 \Gamma(\mu+2)}}{2\pi}\sin{\left
(\frac{\pi\mu}{2} \right )}
 \nonumber\\
& \times\delta(Z) \int {d^2 q}
 \exp(i\bm q \cdot\bm R_\perp)
 q^{-3-\mu}.
 \label{paircorr2dimensional2}
 \end{align}
The main contribution to the above integral stems from small wavevectors $q$ (of the order of the inverse integral scale of turbulence $L_0$) giving an $\bm R_\perp$-independent constant. Extracting also an $\bm R_\perp$-dependent contribution one obtains that
 \begin{align}
&  \langle n_1(\bm R_\perp,Z)n_1({\bm 0},0)\rangle
=\delta(Z) C_n^2 \LARGE [\mathrm{Const+}
 \nonumber\\
&  \left .+\frac{\Gamma(\mu+2)\Gamma(-1/2-\mu/2)}{2^{\mu+2}\Gamma(3/2+\mu/2)}\sin{\left (\frac{\pi\mu}{2} \right )} R_\perp^{\mu+1}
\right].
 \label{paircorr2delta}
 \end{align}
Note that Eq. \e{paircorr2delta} can be obtained from Eq. \e{paircorr2dimensional2}  using the regularized version of the integral through the replacement of the power law \e{KolmogorovObukhovgeneral} with the corresponding Von K\'arm\'an spectrum \cite{TatarskiiBook1961} obtained by the substitution
$ q^{-3-\mu}\to
 (q^2+q_0^2)^{-3/2-\mu/2},$ where $q_0\sim 1/L_0.$

\subsection{Spatial scales}
\label{sec:Spatialscales}

The basic equation (\ref{nlsdimensionall}) is a linear equation with multiplicative noise. There are some characteristic distances for the laser beam propagation in the random medium which can be extracted by comparison of different terms in Eq. (\ref{nlsdimensionall}) and taking into account the expression (\ref{paircorr2delta}). The first scale is determined by the propagation distance at which the scintillation index $\sigma^2_I$ of the initially plane wave, $\Psi|_{Z=0}\equiv const$, becomes $\sim 1$. Using the perturbation technique of Refs. \cite{TatarskiiBook1971,RytovKravtsovTatarskiiBook1989} we obtain that
\begin{align}\label{sigmaR}
\sigma_I^2\equiv \sigma_R^2=p_1C_n^2k_0^{3/2-\mu/2}Z^{3/2+\mu/2}.
\end{align}
Here
\begin{equation}\label{p1def}
p_1\equiv \sin \left(\frac{1}{2} \pi\mu\right) \cos \left(\frac{\pi  (\mu+1)}{4}\right) \Gamma \left(-\frac{\mu}{2}-\frac{3}{2}\right) \Gamma (\mu+2),
\end{equation}
which is the generalization of Eq. (47.31)  of Ref. \cite{TatarskiiBook1971} beyond the particular case $\mu=2/3$. The quantity $\sigma_R^2$ is called the Rytov variance and for $\mu=2/3$ it recovers the standard expression
 \begin{align}\label{sigmaR5p3}
 \sigma_R^2=\frac{\sqrt{3}}{4} \sqrt{2- \sqrt{3}}
 \Gamma \left(-\frac{11}{6}\right)
 \Gamma \left(\frac{8}{3}\right)C_n^2k_0^{7/6}Z^{11/6}
 \nonumber\\
 =1.22871\ldots C_n^2k_0^{7/6}Z^{11/6}
 \end{align}
(see e.g. Ref. \cite{AndrewsPhillipsBook1998}, where $1.22871\ldots $ was replaced by $1.23$ following the approximate numerical value provided in Refs. \cite{TatarskiiBook1971,RytovKravtsovTatarskiiBook1989}).

Fluctuations of the intensity $I$ are strong at $\sigma^2_R\gtrsim 1$, so we define the characteristic distance $Z_{rytov}$ from the condition that $\sigma_R^2=1$, which gives together with Eq. \e{sigmaR}
 \begin{equation}
 Z_{rytov}=\left (p_1C_n^2k_0^{3/2-\mu/2}\right)^{-2/(3+\mu)}.
 \label{rytov}
 \end{equation}
We call the distance (\ref{rytov}) as the {\it\ Rytov length}. In our work the Rytov length is assumed to be larger than the Rayleigh length (\ref{Rayleighlength}),
 \begin{equation}
 Z_{rytov}\gtrsim Z_R,
 \label{narrow}
 \end{equation}
which implies a smallness of the initial beam waist $w_0$ and a relative weakness of the refractive index fluctuations to make sure that the condition \e{narrow} is satisfied. We investigate the propagation distances $Z>Z_{rytov}$ where the Gaussian beam is already disintegrated into speckles.

It follows from Eq. (\ref{nlsdimensionall}) that the envelop $\Psi$ changes with $Z$ due to $n_1$ as $\Psi\propto \exp[ i k_0\int _0^ZdZ' n_1(\bm R_\perp,Z')]$. Then we obtain from Eq.   (\ref{paircorr2delta}) that the average square of the phase difference between two points at the same  $Z$ but different $\bm R_\perp$ caused by the refractive index fluctuations is
 \begin{eqnarray}
 k_0^2 \left\langle\left\{\int \limits_0^ZdZ\, '[n_1(\bm R_\perp,Z')-n_1(\bm 0,Z')]\right\}^2 \right\rangle
 \nonumber \\
 =-\frac{k_0^2\Gamma(\mu+2)\Gamma(-1/2-\mu/2)}{2^{\mu+1}\Gamma(3/2+\mu/2)}\sin{\left (\frac{\pi\mu}{2} \right )}C_n^2 R_\perp^{\mu+1} Z.
 \nonumber
 \end{eqnarray}
Equating this phase difference to unity, we find the phase correlation length
 \begin{equation}
 R_{ph}\sim (k_0^2 C_n^2 Z)^{-1/(\mu+1)}.
 \label{phasecorrelation}
 \end{equation}
Thus qualitatively the beam front at a given $Z$ can be considered as multiple cells of the transverse size $R_{ph}$ with independent phases between different cells.

In accordance with the Huygens-Fresnel principle, the wave amplitude can be considered as a superposition of waves emitted by the secondary sources at a wavefront. The source of a transverse size $R_0$ produces the beam of the transverse length $\sim \theta Z $ at the propagation distance $Z$, where $\theta \equiv 1 /(k_0 R_{0})$ is the corresponding beam divergence. Respectively, the sources of the transverse size $\sim R_{ph}$ located at $Z=Z_1$ produce the beams of the transverse length $\sim  Z _1/(k_0 R_{ph})$ at the distance $Z=2Z_1$. This length becomes larger than the transverse size $2Z_1/(k_0 w_0)$ at $Z=2Z_1$ of the initial beam if $R_{ph}\gtrsim w_0/2$. This condition is satisfied at the distance $Z> Z_\star, $ where  $Z_\star$ is estimated from the condition that  $w_0\sim R_{ph}|_{Z=Z_\star}$ and Eq. \e{phasecorrelation} as
 \begin{equation}
 Z_\star \sim k_0^{-2} C_n^{-2} w_0^{-\mu-1}.
 \label{Andrews}
 \end{equation}
At $Z>Z_\star$ the total transverse beam width $R_{width}$ looses memory of the initial beam waist $w_0$ and is estimated as
 \begin{equation}
 R_{width}\sim Z /(k_0 R_{ph})
 \sim (C_n^2 Z^{\mu+2}k_0^{1-\mu})^{1/(\mu+1)}.
 \label{beamwidth}
 \end{equation}
Note that the inequality (\ref{narrow}) results in $Z_\star\gtrsim Z_{rytov}$.

\section{Dimensionless variables}
\label{sec:dimensionless}

Here we introduce dimensionless parameters, that we use below. Namely, the dimensionless coordinates $\bm r=(x,y)$ and $z$ are defined as follows
 \begin{equation}\label{dimensionless}
 x=X/w_0, \, y= Y/w_0, \, \bm r= \bm R_\perp/ w_0, \, z=Z/(4Z_R).
 \end{equation}
where $w_0$ is the initial Gaussian beam waist, see Eq. (\ref{Gaussianbeam}), and $Z_R$ is Rayleigh length (\ref{Rayleighlength}). Then we obtain from Eq. \e{nlsdimensionall} the following dimensionless stochastic equation
\begin{equation} \label{laser1}
\I\frac{\partial }{\partial  z}\Psi+\nabla^2\Psi
+\xi ({\bm r},z)\Psi=0,
\end{equation}
where the random factor
\begin{equation} \label{nxirelation}
\xi=2k_0 ^2w_0^2n_1,
\end{equation}
determines stochastic properties of the envelope $\Psi$. The initial condition (\ref{Gaussianbeam}) in the dimensionless units  takes the following form
 \begin{equation}
 \Psi_\mathrm{in}(\bm r)= \exp(-r^2),
 \label{initial}
 \end{equation}
where the initial beam amplitude is set to one without loss of the generality because we consider the linear equation for the wave amplitude.

In the dimensionless units the relation (\ref{paircorr2delta}) is rewritten as
 \begin{eqnarray}
 \langle \xi(\bm r_1,z_1)\xi(\bm r_2,\bm z_2)\rangle
 =\left({\rm const}- D r_{12}^{\mu+1}\right)
 \delta(z_1-z_2),
 \label{noise}
 \end{eqnarray}
where $r_{12}=|\bm r_1-\bm r_2|$ and the factor $D$ is
 \begin{eqnarray}
 D=- \frac{ c_n^2 }{2^{\mu+1}}
 \frac{\Gamma(\mu+2)\Gamma(-1/2-\mu/2)}{\Gamma(3/2+\mu/2)}{}\sin{\left
(\frac{\pi\mu}{2} \right )}.
 \label{Ddef}
 \end{eqnarray}
Here we used the dimensionless turbulence strength $c^2_n$ first introduced in Ref. \cite{LushnikovVladimirovaJETP2018} as
\begin{equation}\label{cndef}
\begin{split}
c_n^2\equiv k_0^3w_0^{\mu+3}C_n^2.
\end{split}
\end{equation}
For $\mu=2/3,$  Eq. \e{Ddef} implies that $D/c_n^2=2.91438\ldots$

\subsection{Reduction of the refractive index fluctuations to Gaussian stochastic process with power law pair correlation function}
\label{sec:Gaussianfluctuations}

We introduce the Fourier transform
 \begin{equation*}
 \tilde\xi(\bm k, z)=\int d^2r\ \exp(-i\bm k \bm r) \xi(\bm r,z).
 \end{equation*}
Then Eq. (\ref{noise}) implies that
 \begin{eqnarray}
 \langle \tilde\xi(\bm k,z_1) \tilde \xi(\bm q,z_2) \rangle
 =\frac{D(2\pi)^2}{p_0 k^{3+\mu}}\delta(z_1-z_2)\delta(\bm k+\bm q), \nonumber \\
p_0\equiv\frac{1}{2^{2+\mu}\pi (\mu+1)}
 \frac{\Gamma(1/2-\mu/2)}{\Gamma(3/2+\mu/2)}.
 \label{spectrum}
 \end{eqnarray}
Of course, the expression (\ref{spectrum}) corresponds to Eq. (\ref{noise}).

Since $\xi$ is short-correlated in $z$, it can be regarded to possess Gaussian statistics by the central limit theorem (see e.g. \cite{FellerBook1957}). The probability distribution describing fluctuations of $\xi$ corresponding to Eq. (\ref{spectrum}) can be written as
 \begin{eqnarray}
 {\cal P}\propto \exp\left[ -\frac{p_0}{2D}
 \int dz\ \int \frac {d^2 q}{(2\pi)^2} q^{3+\mu} |\tilde\xi|^2\right],
 \label{probab}
 \end{eqnarray}
where the integration over $z$ is taken over the propagation length of the laser beam. The consistency of Eqs. (\ref{spectrum}) and \e{probab} can be immediately verified by replacing the integrals in Eq. \e{probab} by discrete sums with Gaussian integrals explicitly evaluated which allow to take the continuous limit back from sum to integrals. The probability density \e{probab} can be a starting point for calculating complicated averages over the $\xi$-statistics.

\subsection{Parameters and regions}
\label{sec:Parametersregions}

In the dimensionless variables \e{dimensionless}, using Eqs. \e{p1def}, \e{cndef} and \e{Ddef}, the Rytov length \e{rytov} takes the following form
 \begin{align}
 z_{rytov}\equiv \frac{Z_{rytov}}{4Z_R}
 =\frac{1}{2}  (p_1{c_n^2})^{-\frac{2}{\mu+3}}=p_2D^{-\frac{2}{\mu+3}},
 \label{rytovdimensionless}
 \end{align}
where
\begin{align}\label{p2def}
p_2\equiv\frac{1}{2} \Gamma \left(\frac{  \text{} 2^{{\mu}+3} \pi\cos \left[\frac{1}{4} \pi  (\mu+1)\right] }{(\mu+3)^2 \Gamma \left(-\frac{\mu}{2}-\frac{3}{2}\right)\cos \left(\frac{\pi  \mu}{2}\right)}\right)^{-\frac{2}{\mu+3}}.
\end{align}
For $\mu=2/3$, Eq. \e{p2def} reduces to $p_2=0.80088\ldots.$ The condition \e{narrow} together with \e{rytov} can be rewritten as $D \lesssim 1$.

In the dimensionless variables \e{dimensionless}, the variations of the phase in the transverse direction $\bm r$ due to the noise become of order unity at the scale $r\sim r_{ph}$ where
 \begin{equation}
 r_{ph}= R_{ph}/w_0\sim (Dz)^{-1/(\mu+1)},
 \label{phasel}
 \end{equation}
and $R_{ph}$ is defined by Eq. (\ref{beamwidth}). This quantity can be treated as the phase correlation length of the envelope $\Psi$ in the transverse direction.

Eq. (\ref{Andrews}), with Eqs. \e{Ddef} and \e{cndef} taken into account, transforms in the dimensionless variables \e{dimensionless} into
 \begin{equation}
 z_\star=Z_\star/(4Z_R)\sim D^{-1}.
 \label{star}
 \end{equation}
Note that $z_\star\gtrsim z_{rytov}$ because of the inequality (\ref{narrow}). As discussed in section \ref{sec:Spatialscales}, the total transverse beam width $r_{width}$  at the distance $z\gtrsim z_\star$ is determined by the random diffraction, where $r_{width}$ is determined from Eq. \e{beamwidth} as
 \begin{equation}
 r_{width}=R_{width}/w_0\sim(Dz)^{1/(\mu+1)}z.
 \label{tord8}
 \end{equation}
The width (\ref{tord8}) grows as $z$ increases faster than  the pure diffraction case since $\mu<1$.

The random diffraction leads to a random phase of the field $\Psi$ at $z\gg z_\star$. Therefore $\Psi$ possesses Gaussian statistics at the scales. Correspondingly, the intensity $I=|\Psi|^2$ has the Poissonian statistics. We examine below the accuracy of this statement by calculating corrections to the Poissonian statistics.

\section{Correlation functions}
\label{sec:corr}

Statistical properties of the electric field envelope $\Psi$ can be examined in terms of its correlation functions. The second-order correlation function is the average
 \begin{equation}
 F_2(\bm r_1,\bm r_2,z) =\left\langle \Psi(\bm r_1, z) \Psi^\star(\bm r_2, z)\right\rangle,
 \label{per2}
 \end{equation}
taken at a given $z$. We examine also higher order correlation functions
 \begin{eqnarray}
 F_{2n}(\bm r_1, \dots , \bm r_{2n},z) = \qquad
 \nonumber \\
 \langle \Psi(\bm r_1,z) \dots
 \Psi(\bm r_n,z) \Psi^\star(\bm r_{n+1},z) \dots
 \Psi^\star(\bm r_{2n},z) \rangle,
 \label{cof1}
 \end{eqnarray}
The correlation functions (\ref{per2},\ref{cof1}) are obviously invariant under a homogeneous phase shift. Therefore they are insensitive to the refraction index fluctuations at the integral scale of turbulence.

The angular brackets in Eqs. (\ref{per2}) and (\ref{cof1}) designate averaging over the statistics of $\xi$. Principally, one should solve Eq. (\ref{laser1}) for any realization $\xi$ at a given initial condition, then calculate the product in the angular brackets in Eq. (\ref{cof1}) and then average over the realizations with the weight dictated by Eq. (\ref{noise}). Of course, this procedure cannot be performed explicitly. However, Eq. (\ref{laser1}) and the expression (\ref{noise}) admit derivation of closed equations for correlation functions $F_{2n}$. First such procedure for the pair correlation function was proposed by Kraichnan \cite{Kraichnan} and Kazantsev \cite{Kazantsev} in the contexts of the passive scalar turbulence and turbulent dynamo, correspondingly. It was independently obtained in the optical context in Ref. \cite{TatarskiiJETP1969} for both the pair and higher order correlation functions.

To obtain the equations for $F_{2n}$, one may start with the relation
 \begin{equation}
 \Psi(\bm r, z_2)=\mathrm T\exp\left[i\int_{z_1}^{z_2}dz\ (\nabla^2+\xi)\right]\Psi(\bm r,z_1),
 \label{tord}
 \end{equation}
where $\mathrm T\exp$ means an $z$-ordered exponent. The relation (\ref{tord}) is a direct consequence of the equation (\ref{laser1}). The equation (\ref{tord}) enables one to relate a product of $\Psi(z_2)$, $\Psi^\star(z_2)$ to the corresponding product of $\Psi(z_1)$, $\Psi^\star(z_1)$. Due to short in $z$ correlations of $\xi$ one can independently average the average of $\Psi(z_1)$, $\Psi^\star(z_1)$ and the exponents. Say,
 \begin{eqnarray}
 F_2(\bm r_1,\bm r_2,z_2) =\left\langle \Psi(\bm r_1, z_2) \Psi^\star(\bm r_2, z_2)\right\rangle
 \nonumber \\
 =\left\langle \mathrm T\exp\left\{i\int_{z_1}^{z_2}dz\ [\nabla_1^2+\xi(\bm r_1)]\right\} \right.
 \nonumber \\
 \left. \mathrm T\exp\left\{-i\int_{z_1}^{z_2}dz\ [\nabla_2^2+\xi(\bm r_2)]\right\} \right\rangle
 \nonumber \\
 F_2(\bm r_1,\bm r_2,z_1).
 \label{tord1}
 \end{eqnarray}
Analogous relations can be obtained for other products.

Analyzing close $z_1$ and $z_2$, one can expand the exponents. Since $\langle \xi \rangle=0$, we should expand terms with $\xi$ up to the second order as follows
 \begin{eqnarray}
 \mathrm T\exp\left[i\int_{z_1}^{z_2}dz\ (\nabla^2+\xi)\right]\approx
 1+i\int_{z_1}^{z_2}dz\ (\nabla^2+\xi)
 \nonumber \\
 -\int_{z_1}^{z_2}dz\ \xi(\bm r, z) \int_{z_1}^z d\zeta\ \xi(\bm r,\zeta).
 \label{tord2}
 \end{eqnarray}
Substituting the expansion (\ref{tord2}) and the analogous expressions for the other exponents, keeping terms up to the second order in $\xi$ and averaging, we relate $F_2(z_2)$ to $F_2(z_1)$. Since $\langle\xi(\bm r,z_1)\xi(\bm r,z_2)\rangle$ is a narrow symmetric function of $z_1-z_2$,
one should take
 \begin{equation*}
 \int^{z_2} dz_1\, \langle\xi(\bm r,z_1)\xi(\bm r,z_2)\rangle=\mathrm{const}/2,
 \end{equation*}
see Eq. (\ref{noise}). As a result, we obtain an increment of $F_2$, proportional to $z_2-z_1$. Passing from the (small) increment to the differential equation, one finds the equation for the pair correlation function (\ref{per2})
 \begin{eqnarray}
 \partial_z F_2=i(\nabla_1^2-\nabla_2^2)F_2
 -D r^{\mu+1} F_2,
 \label{per1}
 \end{eqnarray}
where $\bm r=\bm r_1-\bm r_2$. The constant, which appears in Eq. (\ref{noise}), drops from the equation, as it should be. The equation (\ref{per1}) can be rewritten as
 \begin{equation}
 \partial_z F_2=2i\frac{\partial^2}{\partial\bm R \partial\bm r} F_2
 -D r^{\mu+1} F_2,
 \label{per3}
 \end{equation}
where $\bm R=(\bm r_1+\bm r_2)/2$.

\subsection{Pair correlation function}

A formal solution of the equation (\ref{per1}) can be written in terms of the two-point Green function ${\cal G}$:
 \begin{eqnarray}
 F_2(\bm r_1,\bm r_2,z)=
 \int d^2x_1\, d^2x_2\, {\cal G}(\bm r_1,\bm r_2,\bm x_1,\bm x_2,z)
 \nonumber \\
 \Psi_{in}(\bm x_1) \Psi_{in}(\bm x_2),
 \label{tord3}
 \end{eqnarray}
where $\Psi_{in}(\bm x_1) \Psi_{in}(\bm x_2)$ is the initial value of the pair correlation function, see Eq. (\ref{initial}). The Green function ${\cal G}$ is equal to zero at $z<0$ and satisfies the equation
 \begin{eqnarray}
 \partial_z {\cal G}=i(\nabla_1^2-\nabla_2^2){\cal G}
 -D r^{\mu+1} {\cal G}
 \nonumber \\
 +\delta(z) \delta(\bm r_1-\bm x_1) \delta(\bm r_2-\bm x_2).
 \label{tord4}
 \end{eqnarray}
Note that the Green function by itself does not know about initial conditions for the envelope $\Psi$.

To find the Green function, one can pass to the Fourier transform $\tilde {\cal G}$ as follows
 \begin{equation}
 {\cal G}(\bm R,\bm r,\bm X,\bm x,z)=\int \frac{d^2 k}{(2\pi)^2}
 \exp(i\bm k \bm R) \tilde {\cal G}(\bm k,\bm r, \bm X, \bm x,z),
 \nonumber
 \end{equation}
where $\bm r=\bm r_1-\bm r_2$, $\bm R=(\bm r_1+\bm r_2)/2$, $\bm x=\bm x_1-\bm x_2$, $\bm X=(\bm x_1+\bm x_2)/2$. Then the equation (\ref{tord4}) is rewritten as
 \begin{eqnarray}
 \partial_z \tilde {\cal G}=-2\bm k \nabla \tilde {\cal G} -D r^{\mu+1} \tilde {\cal G}
 \nonumber \\
 =\delta(z) \delta(\bm r-\bm x) \exp(-i\bm X \bm k),
 \nonumber
 \end{eqnarray}
where $\nabla=\partial/\partial\bm r$. Solving that equation by characteristics, one finds
 that \begin{eqnarray}
 \tilde {\cal G}=\theta(z) \delta(\bm r-2\bm kz-\bm x) \exp(-i\bm X \bm k)
 \nonumber \\
 \exp\left[-D\int_0^z d\zeta\ |\bm r-2 \bm k \zeta|^{\mu+1}\right].
 \label{per4}
 \end{eqnarray}
Returning to the real space, one finds
 \begin{eqnarray}
 {\cal G}= \frac{\theta(z)}{16 \pi^2 z^2}
 \exp\left[\frac{i}{2z}(\bm r-\bm x)(\bm R-\bm X) \right.
 \nonumber \\
 \left.
 -{Dz}\int_0^1 d\chi\, |\chi \bm x+(1-\chi)\bm r|^{\mu+1}  \right].
 \label{tord5}
 \end{eqnarray}
This quantity is symmetric in permutation of the initial and terminal points, as it should be.

Integrating the expression (\ref{tord5}) over $\bm R$ at a given $\bm r$, one obtains
that  \begin{eqnarray}
 \int d^2R\ {\cal G}= {\theta(z)}\delta(\bm r-\bm x)
 \exp(-Dzr^{\mu+1}).
 \label{tord9}
 \end{eqnarray}
Substituting the expression into Eq. (\ref{tord3}), we find that
 \begin{equation}
 \int d^2R\ F_2=\frac{\pi}{2}
 \exp\left(-\frac{r^2}{2}-Dzr^{\mu+1}\right),
 \label{tord10}
 \end{equation}
where the expression (\ref{initial}) is substituted. The first term in the exponent dominates at $z\ll z_\star$ and the second term dominates in the opposite limit, $z\gg z_\star$. The dominance of the term $-Dzr^{\mu+1}$ in the second regime  $z\gg z_\star$ implies that the transverse correlation length $r_{ph}\sim (Dz)^{-1/(\mu+1)}$ in full agrement with the qualitative analysis of sections \ref{sec:Spatialscales} and \ref{sec:Parametersregions} including Eq. \e{phasel}.

Further we are interested in distances $z\gg z_\star$, where effects of random diffraction are relevant. Then the characteristic values of $\bm x$ and $\bm r$ are determined by the integral in the exponent in Eq. (\ref{tord5}). Equating the integral to unity, we find the estimate $|\bm x|\sim |\bm r|\sim r_{ph}$, where $r_{ph}$ is the phase correlation length (\ref{phasel}). The correlation length is related to the random diffraction on fluctuations of the refraction index destroying phase correlations. Equating then the first term in the exponent in Eq. (\ref{tord5}), we find the characteristic value $R\sim z/r_{ph}=r_{width}$ where $r_{width}$ is determined by Eq. (\ref{tord8}). The quantity has the meaning of the beam width, caused by the random diffraction. At distances $z\gg z_\star$ the width is much larger than the pure diffraction width $z$. In analyzing the pair correlation function in accordance with Eq. (\ref{tord3}), one should take into account the initial width of the beam. Just the initial width determines the characteristic value of $\bm X$, it can be estimated as unity.

Thus in the case $z\gg z_\star$ one obtains from Eq. (\ref{tord3})
 \begin{eqnarray}
  \langle \Psi(\bm R+\bm r/2) \Psi^\star(\bm R-\bm r/2) \rangle
  =\int \frac{d^2x\, d^2X}{16 \pi^2 z^2}
  \nonumber \\
 \exp\left[\frac{i}{2z}\bm r\bm R
 -{Dz}\int_0^1 d\chi\, |\chi \bm r+(1-\chi)\bm x|^{\mu+1}  \right]
 \nonumber \\
 \Psi_{in}(\bm X+\bm x/2)\Psi_{in}(\bm X-\bm x/2),
 \label{pairc3}
 \end{eqnarray}
where we neglected $\bm X$, $\bm x$ in the first term in the exponent. If $r\ll r_{ph}$ then the value of $x$ is determined by the second term in the exponent, then $x$ can be estimated as $r_{ph}$. If $r_{ph}\ll r \ll 1$ then the value of $x$ is determined by the second term in the exponent as well, however, $x$ can be estimated as $r$. After integration over $\bm x$ in the expression (\ref{pairc3}), there remains a dependence on $\bm r$ with the characteristic value $r\sim r_{ph}$. Since at $r\ll 1$, $x\ll 1$ as well, we can neglect $\bm x$ in the product $\Psi_{in}(\bm X+\bm x/2)\Psi_{in}(\bm X-\bm x/2)$ in Eq. (\ref{pairc3}). Then one can integrate over $\bm X$ to obtain
 \begin{eqnarray}
  \langle \Psi(\bm R+\bm r/2) \Psi^\star(\bm R-\bm r/2) \rangle
  = \int \frac{d^2x}{32 \pi z^2}
  \nonumber \\
 \exp\left[\frac{i}{2z}\bm r\bm R
 -{Dz}\int_0^1 d\chi\, |\chi \bm r+(1-\chi)\bm x|^{\mu+1}  \right].
 \label{pairc4}
 \end{eqnarray}

Analyzing the expression (\ref{pairc4}) we conclude that the characteristic value of $r$ is determined by the expression (\ref{phasel}). Thus, the quantity $r_{ph}$ plays the role of the beam
correlation length in the transverse direction as well. We neglected the factor $\exp [-{i}\bm x\bm R/(2z)]$ in the expression (\ref{pairc4}). For $R$ larger than $z/R_{ph}=D^{1/{\mu+1}}z^{1/{\mu+1}+1}$, the exponent is fast oscillating. That leads to diminishing the expression of the pair correlation function in comparison with the expression (\ref{phasel}).
Thus, the quantity (\ref{tord8}) is the beam width for $z\gg z_\star$, indeed. The quantity is determined solely by fluctuations.

\subsection{Fourth-order correlation function}

The equation for the fourth-order correlation function
 \begin{equation}
 F_4=\langle \Psi(\bm r_1,z) \Psi(\bm r_2,z)
 \Psi^\star(\bm r_3,z) \Psi^\star(\bm r_4,z) \rangle
 \label{per9}
 \end{equation}
can be derived, similar the equation (\ref{per1}), from the representation (\ref{tord}). The corresponding equation is give by
 \begin{eqnarray}
 \partial_z F_4= i(\nabla_1^2+\nabla_2^2-\nabla_3^2-\nabla_4^2)F_4 \qquad
 \label{per10} \\
 -D\left[-r_{12}^{\mu+1}+r_{13}^{\mu+1}+r_{14}^{\mu+1}-r_{34}^{\mu+1}
 +r_{23}^{\mu+1}+r_{24}^{\mu+1}\right]F_4,
 \nonumber
 \end{eqnarray}
where $r_{12}=|\bm r_1-\bm r_2|$ and so on.

Generally, the separations between the points in the different spots are of the order of $r_{width}$ (\ref{tord8}). Then the real factor in the right-hand side of the equation (\ref{per10}) is $\sim Dr_{width}^{\mu+1}$. At $z\gg z_\star$ one finds $Dr_{width}^{\mu+1}z\gg1$. That leads to a strong suppression of the fourth-order correlation function. However, in the geometry where separations between the points $\bm r_1$ and $\bm r_3$, $\bm r_2$ and $\bm r_4$ (or  between the points $\bm r_1$ and $\bm r_4$, $\bm r_2$ and $\bm r_3$) are much smaller than $r_{width}$, the factor appears to be much smaller. Therefore the fourth-order correlation function has sharp maxima in the geometries. Further we examine just this case.

Having in mind the geometry where separations $r_{13}$ and $r_{24}$ are much smaller than $r_{width}$, we rewrite the equation (\ref{per10}) as
 \begin{eqnarray}
 \frac{\partial F_4}{\partial z}=2i\left(\frac{\partial^2F_4}{\partial\bm R_1 \partial \bm \rho_1}
 +\frac{\partial^2F_4}{\partial\bm R_2 \partial \bm \rho_2}\right) \qquad
 \nonumber \\
 - [D(\rho_1^{\mu+1}+\rho_2^{\mu+1})+U] F_4,  \qquad
 \label{perr} \\
 U/D =|\bm R+\bm\rho_1/2+\bm\rho_2/2|^{\mu+1}
 -|\bm R+\bm\rho_1/2-\bm\rho_2/2|^{\mu+1}
\nonumber \\
 -|\bm R-\bm\rho_1/2+\bm\rho_2/2|^{\mu+1}
 +|\bm R-\bm\rho_1/2-\bm\rho_2/2|^{\mu+1}, \quad
 \label{peru}
 \end{eqnarray}
where we introduced
 \begin{eqnarray}
 \bm r_1=\bm R_1+\bm\rho_1/2, \
 \bm r_3=\bm R_1-\bm\rho_1/2,
 \nonumber \\
 \bm r_2=\bm R_2+\bm\rho_2/2, \
 \bm r_4=\bm R_2-\bm\rho_2/2,
 \nonumber
 \end{eqnarray}
and $\bm R=\bm R_1-\bm R_2$. If $R\sim R_{width}\gg \rho_1,\rho_2$. Then the terms in the quantity $U$ (\ref{peru}) cancel each other and we can neglect $U$ in the equation (\ref{perr}). Then the operator in the equation becomes a sum of two operators in the equation for the pair correlation function (\ref{tord4}). Therefore in the geometry, $F_4$ is a product of two pair correlation functions, $F_2(\bm r_1,\bm r_3)F_2(\bm r_2,\bm r_4)$.

Let us now estimate accuracy of the approximation. For the purpose we should evaluate corrections to the fourth-order correlation function $F_4$. In accordance with the equation (\ref{perr}), the first correction is written as
 \begin{eqnarray}
 \delta F_4= -\int_0^z d\zeta \int d^2x_1 d^2x_2d^2x_3d^2x_4
 \nonumber \\
 U(\bm x_1, \bm x_2,\bm x_3,\bm x_4)
 F_2(\bm x_1,\bm x_3,\zeta) F_2(\bm x_2,\bm x_4,\zeta)
 \nonumber \\
 {\cal G}(\bm r_1,\bm r_3,\bm x_1,\bm x_3,z-\zeta)
 {\cal G}(\bm r_2,\bm r_4,\bm x_2,\bm x_4,z-\zeta).
 \label{peru2}
 \end{eqnarray}
In this integral the separations $x_{13}\sim x_{24}\sim r_{ph}$, $x_{14}\approx x_{23}\sim r_{width}$, and the quantity
 \begin{equation*}
 U\sim D x_{14}^{c-2}x_{13}^2\sim D r_{ph}^2 r_{width}^{c-2}.
 \end{equation*}
Evaluating the Green functions ${\cal G}$ in Eq. (\ref{peru2}) as $z^{-2}$ in accordance with the expression (\ref{tord5}), we find that the correction (\ref{peru2}) is evaluated as $\alpha F_2(\bm r_1,\bm r_3)F_2(\bm r_2,\bm r_4)$, where $\alpha$ is the parameter
 \begin{equation}
 \alpha=(z_{rytov}/z)^{(1-\mu)(\mu+3)/(\mu+1)}.
 \label{smallpar}
 \end{equation}
The parameter is small provided $z\gg z_{rytov}$. A more accurate calculation is presented in Appendix \ref{sec:crossbar}.

To analyze the fourth-order correlation function $F_4$ in the geometry, where all separations between the points are much smaller than $r_{width}$, one should use the Green function
 \begin{equation*}
{\cal G}_4(\bm r_1,\bm r_2,\bm r_3,\bm r_4,\bm x_1,\bm x_2,\bm x_3,\bm x_4,z),
 \end{equation*}
of the equation (\ref{per10}), satisfying
 \begin{eqnarray}
 \partial_z {\cal G}_4- i(\nabla_1^2+\nabla_2^2-\nabla_3^2-\nabla_4^2){\cal G}_4+
 \nonumber \\
 D\left[-r_{12}^{\mu+1}+r_{13}^{\mu+1}+r_{14}^{\mu+1}-
 r_{34}^{\mu+1}+r_{23}^{\mu+1}+r_{24}^{\mu+1}\right]{\cal G}_4
 \nonumber \\
 =\delta(z) \delta(\bm r_1-\bm x_1)\delta(\bm r_2-\bm x_2)
 \delta(\bm r_3-\bm x_3)\delta(\bm r_4-\bm x_4).
 \label{peru3}
 \end{eqnarray}
The fourth-order correlation function is expressed as
 \begin{eqnarray}
 F_4(\bm r_1,\bm r_2,\bm r_3,\bm r_4,z)
 =\int d^2 x_1 d^2 x_2d^2x_3d^2x_4
 \nonumber \\
 {\cal G}_4(\bm r_1,\bm r_2,\bm r_3,\bm r_4,\bm x_1,\bm x_2,\bm x_3,\bm x_4,z)
 \nonumber \\
 \Psi_{in}(\bm x_1) \Psi_{in}(\bm x_2) \Psi_{in}(\bm x_3)\Psi_{in}(\bm x_4).
 \label{peru4}
 \end{eqnarray}

One can apply the same arguments as for the fourth order correlation function, to the Green function ${\cal G}$. Thus, ${\cal G}$ as a function of $\bm r_1,\bm r_2,\bm r_3,\bm r_4$ has the sharp maxima in the geometry where separations between the points $\bm r_1$ and $\bm r_3$, $\bm r_2$ and $\bm r_4$ (or  between the points $\bm r_1$ and $\bm r_4$, $\bm r_2$ and $\bm r_3$) are much smaller than $r_{width}$. Therefore in the main approximation
 \begin{eqnarray}
 {\cal G}_4(\bm r_1,\bm r_2,\bm r_3,\bm r_4,\bm x_1,\bm x_2,\bm x_3,\bm x_4,z) \approx
 \nonumber \\
 {\cal G}(\bm r_1,\bm r_3,\bm x_1,\bm x_3,z)
 {\cal G}(\bm r_2,\bm r_4,\bm x_2,\bm x_4,z)
 \nonumber \\
 {\cal G}(\bm r_1,\bm r_4,\bm x_1,\bm x_4,z)
 {\cal G}(\bm r_2,\bm r_3,\bm x_2,\bm x_3,z) .
 \label{peru5}
 \end{eqnarray}
Since the equation for the correlation function ${\cal G}_4$ in terms of $\bm x_i$ is the same as in terms of $\bm r_i$ (\ref{peru3}), the approximation (\ref{peru5}) is correct in terms of $\bm x_i$ as well.

As for any Green function, one may write
 \begin{eqnarray}
 {\cal G}_4(\bm r_1,\bm r_2,\bm r_3,\bm r_4,\bm x_1,\bm x_2,\bm x_3,\bm x_4,z) =
 \nonumber \\
 \int d^2y_1 d^2y_2 d^2y_3d^2y_4
 \nonumber \\
 {\cal G}_4(\bm r_1,\bm r_2,\bm r_3,\bm r_4,\bm y_1,\bm y_2,\bm y_3,\bm y_4,\zeta)
 \nonumber \\
 {\cal G}_4(\bm y_1,\bm y_2,\bm y_3,\bm y_4,\bm x_1,\bm x_2,\bm x_3,\bm x_4,z-\zeta). \qquad
 \label{peru6}
 \end{eqnarray}
In integration over $\bm y_i$, the main contribution to the integral is produced just by the regions where  separations between the points $\bm y_1$ and $\bm y_3$, $\bm y_2$ and $\bm y_4$ (or  between the points $\bm y_1$ and $\bm y_4$, $\bm y_2$ and $\bm y_3$) are much smaller than $r_{width}$. Such integration reproduces the approximation (\ref{peru5}). We conclude that one can use the approximation (\ref{peru5}) for any geometry of the points $\bm r_i,\bm x_i$.

\subsection{Higher order correlation functions}

One can easily generalize the above procedure for correlation functions of arbitrary order. The equation for $2n$-th order correlation function is
 \begin{eqnarray}
 \partial_z F_{2n}=
 i(\nabla_1^2+\dots +\nabla_n^2-
 \nabla_{n+1}^2 - \dots -\nabla_{2n}^2) F_{2n}  \quad
 \label{cof2} \\
 +D \left[ {\sum_{i=1}^n}\sum_{j=i+1}^n r_{ij}^{\mu+1}
 +\sum_{i=n+1}^{2n} \sum_{j=i+1}^{2n} r_{ij}^{\mu+1}
 \right. \nonumber \\ \left.
 -\sum_{i=1}^n \sum_{j=n+1}^{2n}r_{ij}^{\mu+1}
 \right]F_{2n},
 \nonumber
 \end{eqnarray}
where, as above, $r_{ij}=|\bm r_i-\bm r_j|$. Analogously to Eqs. (\ref{tord3},\ref{peru4}), a formal solution of the equation (\ref{cof2}) can be written as
 \begin{eqnarray}
 F_{2n}=\int d^2x_1 \dots d^2x_{2n}
 \nonumber \\
 {\cal G}_{2n}(\bm r_1,\dots, \bm r_{2n},\bm x_1,\dots, \bm x_{2n},z)
 \nonumber \\
 \Psi_{in}(\bm x_1) \dots \Psi_{in}( \bm x_{2n}),
 \label{tord6}
 \end{eqnarray}
where the Green function ${\cal G}_{2n}$ is equal to zero at $z<0$ and satisfies the equation
 \begin{eqnarray}
 \partial_z {\cal G}_{2n}=
 i(\nabla_1^2+\dots +\nabla_n^2-
 \nabla_{n+1}^2 - \dots -\nabla_{2n}^2) {\cal G}_{2n}  \quad
 \nonumber \\
 +D \left[ {\sum_{i=1}^n}\sum_{j=i+1}^n r_{ij}^{\mu+1}
 +\sum_{i=n+1}^{2n} \sum_{j=i+1}^{2n} r_{ij}^{\mu+1}
 \right. \nonumber \\ \left.
 -\sum_{i=1}^n \sum_{j=n+1}^{2n}r_{ij}^{\mu+1}
 \right]{\cal G}_{2n}
 \nonumber \\
 +\delta(z)\delta(\bm r_1-\bm x_1)\dots \delta(\bm r_{2n}-\bm x_{2n}). \qquad
 \label{tord7}
 \end{eqnarray}

Under the condition $z\gg z_\star,$ the Green function ${\cal G}_{2n}$ have sharp maxima in the geometry where the points are split into $n$ close pairs $\bm r_i,\bm r_j$ ($i=1,\dots, n$, $j=n+1,\dots,2n$), where the separations are smaller or of the order of $r_{ph}$ (\ref{phasel}). In the geometry, the ``large'' differences (of the order of $r_{width}$) in Eq. (\ref{tord7}) cancel each other. Then we pass to the differential operator that is a sum of the differential operators of the type appearing in the equation for the pair correlation function (\ref{per1}). Therefore the $2n$-th order correlation function in the geometry is a product of $n$ pair correlation functions. There are $n!$ of such geometries, and in the main approximation the $2n$-th order correlation function can be presented as a sum of $n!$ terms that are products of the pair correlation functions, like the expression (\ref{peru6}). This is just the case subjected by Wick theorem where $\Psi$ possesses Gaussian statistics. The property is a consequence of random diffraction that makes the phase of $\Psi$ a random variable.

One can estimate corrections to the Gaussian statistics. For this one should evaluate the terms that were discarded in the geometry of close pairs. An analysis, analogous to one produced for the fourth order correlation function, shows that the approximation is justified by the same small parameter $\alpha$ (\ref{smallpar}).

\subsection{Statistics of intensity}

Let us analyze statistical properties of the intensity $I=|\Psi|^2$, taken at the distance $z\gg z_\star$ inside the diffraction spot $r\ll r_{width}$. One expects that the quantity has the Poisson statistics because of the randomness of the phase of $\Psi$ caused by the refractive index fluctuations \cite{Halperin}. We prove the conjecture and give the quantitative criterion determining the applicability region of the statistics.

The average $\langle I^n \rangle$ can be written as
 \begin{eqnarray}
 \langle I^n \rangle=
 \int d^2x_1 \dots d^2x_{2n}
 {\cal G}_{2n}(\bm r,\dots, \bm r,\bm x_1,\dots, \bm x_{2n},z)
 \nonumber \\
 \Psi_{in}(\bm x_1) \dots \Psi_{in}( \bm x_{2n}),
 \qquad
 \label{tord26}
 \end{eqnarray}
in accordance with Eq. (\ref{tord6}). Thus we should establish properties of the Green function ${\cal G}_{2n}$ in the situation where the final points coincide.

We use the following property of any Green function
 \begin{eqnarray}
 {\cal G}_{2n}(\bm r_1,\dots, \bm r_{2n},\bm x_1,\dots, \bm x_{2n},z)
 =\int d^2y_1 \dots d^2y_{2n}
 \nonumber \\
 {\cal G}_{2n}(\bm r_1,\dots, \bm r_{2n},\bm y_1,\dots, \bm y_{2n},\zeta) \qquad
 \nonumber \\
 {\cal G}_{2n}(\bm y_1,\dots, \bm y_{2n},\bm x_1,\dots, \bm x_{2n},z-\zeta) \qquad
 \label{tord27}
 \end{eqnarray}
If we choose $\zeta\sim z,$ then the characteristic value of $y$ in the integral can be estimated as $R_{width}$  (\ref{tord8}). However, both Green functions under the integral have have sharp maxima in the geometry where the points are split into $n$ close pairs $\bm y_i,\bm y_j$ ($i=1,\dots, n$, $j=n+1,\dots,2n$), provided the separations are smaller or of the order of $R_{ph}$ (\ref{phasel}). For the second Green function, we established the property above. For the first Green function it follows from the fact that in terms of the variables $y,$ it satisfies the same equation as in terms of $r$. Thus both Green functions are represented as sums of the products of the pair Green functions. Therefore we arrive at the Gaussian statistics for  $\Psi(\bm r)$ and, consequently, at the Poisson statistics for $I$. In other words, the probability distribution function of $I$ is exponential.

Corrections to the Gaussian approximation are controlled by the same small parameter $\alpha$ (\ref{smallpar}). Now we can estimate the region of applicability of the Poisson approximation. If we analyze $\langle I^n\rangle$, then the relative correction, associated with the neglected terms in the equation for the Green functions, is estimated as $n^2 \alpha$, for large $n$. Thus the Poisson expression is valid if $n \ll 1/\sqrt{\alpha}$. In other words, the exponential probability distribution is correct one if $I\ll \langle I \rangle /\sqrt{\alpha}$.

\section{Effective action}
\label{sec:Effective action}

Here we propose an alternative language for describing effects associated with fluctuations of the refractive index. Correlation functions of the field $\Psi$ can be examined in the framework of an effective quantum field theory \cite{MSR73,Janssen,Domin}. The theory produces a diagrammatic expansion of the type first developed by Wyld in the context of hydrodynamic turbulence \cite{Wyld}. Applications of the technique to the optical problems can be found in the works \cite{Kolokol,Churkin,Vergeles}.

In the framework of the effective quantum theory, the correlation functions of the field $\Psi$ can be found as functional integrals over $\Psi,\Psi^\star,P,P^\star$ (where $P,P^\star$ are auxiliary fields) with the weight
 \begin{equation}
 \exp\left\{-{\cal S}+\int d^2r\ \left[P \Psi_{in}^\star
 +P^\star \Psi_{in}\right]\right\},
 \label{weight}
 \end{equation}
where the effective action ${\cal S}$ is constructed in accordance with the equation (\ref{laser1}). Here $\Psi_{in}$ is the initial condition for the field $\Psi$ posed at $z=0$. In our setup, the initial condition is determined by Eq. (\ref{initial}).

To find ${\cal S}$, we start from the weight
 \begin{equation}
 \exp\left\{-{\cal I}+\int d^2r\ \left[P \Psi_{in}^\star
 +P^\star \Psi_{in}\right]\right\},
 \label{weight1}
 \end{equation}
where ${\cal I}$ forces the equation (\ref{laser1}):
 \begin{eqnarray}
 {\cal I}=\int d^2r\ dz\
 P^\star\left(i\partial_z\Psi + \nabla^2 \Psi +\xi\Psi\right)
 \nonumber \\
 -\int d^2r\ dz\
 P\left(i\partial_z\Psi^\star - \nabla^2 \Psi^\star -\xi\Psi\right).
 \label{weight2}
 \end{eqnarray}
The weight (\ref{weight}) is obtained by averaging over the statistics of the refractive index fluctuations in accordance with Eq. (\ref{noise}). The constant term corresponding to homogeneous phase fluctuations cannot contribute to the effective action. Thus we arrive at the additional condition
 \begin{equation}
 \int d^2r\, (P^\star \Psi+ P\Psi^\star ) =0,
 \label{gauge}
 \end{equation}
to be imposed on the field $P$. Then the constant drops from the consideration.

As a result, we find that the effective action ${\cal S}$ is the sum ${\cal S}= {\cal S}_{(2)} +{\cal S}_{int}$, where
 \begin{eqnarray}
 {\cal S}_{(2)}=\int d^2r\ dz\
 P^\star\left(i\partial_z\Psi + \nabla^2 \Psi \right)
 \nonumber \\
 -\int d^2r\ dz\
 P\left(i\partial_z\Psi^\star - \nabla^2 \Psi^\star \right),
 \label{lader4} \\
 {\cal S}_{int}=\frac{1}{2}\int d^2r_1\, d^2r_2\, dz\
 (P_1^\star \Psi_1+ P_1\Psi^\star_1 )
 \nonumber \\
 (D |\bm r_1-\bm r_2|^{\mu+1}-{\rm const})
 (P_2^\star \Psi_2+ P_2\Psi_2^\star ).
 \label{lasem4}
 \end{eqnarray}
Here the first terms describes the free laser beam diffraction, whereas the second term describes the influence of the fluctuations of the refractive index, that is the random diffraction.

Correlation functions of the fields $\Psi,\Psi^\star,P,P^\star$ can be found in the framework of the perturbation theory.  For the purpose one expands $\exp(-{\cal S}_{int})$ in ${\cal S}_{int}$ and calculate explicitly the resulting expressions that are Gaussian integrals with the weight
 \begin{equation*}
  \exp\left\{-{\cal S}_{(2)}+\int d^2r\ \left[P \Psi_{in}
 +P^\star \Psi_{in}\right]\right\},
 \end{equation*}
that is the exponent of the quadratic in the fields quantity. The terms of the perturbation series can be represented as Feynman diagrams. The diagrammatic expansion of this type was first used by Wyld in the context of hydrodynamic turbulence \cite{Wyld}.

Analytical expressions caused by the diagrams are constructed from the propagators, that are the correlation functions
 \begin{eqnarray}
 G=\langle \Psi(\bm r,z) P^\star(0,0)\rangle =- \frac{\theta(z)}{4\pi z}
 \exp\left(i\frac{r^2}{4z}\right),
 \label{prop1} \\
 G^\star=\langle \Psi^\star(\bm r,z) P(0,0)\rangle =- \frac{\theta(z)}{4\pi z}
 \exp\left(-i\frac{r^2}{4z}\right).
 \label{prop2}
 \end{eqnarray}
Here angular brackets mean averaging with the weight $\exp[-{\cal S}_{(2)}]$ and $\theta(z)$ is Heaviside step function. Let us stress that there are no corrections, caused by the interaction term (\ref{lasem4}), to the expressions (\ref{prop1}) and (\ref{prop2}) because of causality.

\subsection{Ladder approximation}
\label{sec:Ladderapproximation}

The pair correlation function of the envelope at a given $z$ is written as the integral (\ref{tord3}). The Green function in the relation can be expressed in terms of the correlation function of the introduced fields
 \begin{equation}
 {\cal G}=\langle \Psi(\bm r_1,z) \Psi^\star(\bm r_2,z)
 P(\bm x_1,0)P^\star(\bm x_2,0)\rangle.
 \label{dia1}
 \end{equation}
The expression (\ref{dia1}) can be derived if to expand the weight (\ref{weight}) up to the second order in $\Psi_{in}$.

 \begin{figure}
\begin{picture}(200,110)
\put(-20,0){ \begin{tikzpicture}
    \begin{scope}[line width=1.0pt,decoration={
            markings,
            mark=at position 0.53 with {\arrow[scale=1.25,black]{stealth}}},
            ]
    \draw[postaction={decorate}] (-3.5,3.5) -- (-1,3.5);
    \draw[postaction={decorate}](-1,2.5) -- (-3.5,2.5);
    \end{scope}
    \begin{scope}[line width=1.0pt,decoration={
            markings,
            mark=at position 0.53 with {\arrow[scale=1.25,black]{stealth}}},
            ]
    \draw[postaction={decorate}] (0.5,3.5) -- (2.5,3.5);
    \draw[postaction={decorate}](2.5,2.5) -- (0.5,2.5);
    \draw[postaction={decorate}] (2.5,3.5) -- (4.5,3.5);
    \draw[postaction={decorate}](4.5,2.5) -- (2.5,2.5);
    \end{scope}
   \begin{scope}[line width=1.0pt,decoration={
            markings,
            mark=at position 0.53 with {\arrow[scale=1.25,black]{stealth}}},
            ]
    \draw[postaction={decorate}] (-3.5,1) -- (-1,1);
    \draw[postaction={decorate}](-1,1) -- (1.5,1);
    \draw[postaction={decorate}] (1.5,1) -- (4,1);
    \draw[postaction={decorate}](4,0) -- (1.5,0);
    \draw[postaction={decorate}] (1.5,0) -- (-1,0);
    \draw[postaction={decorate}](-1,0) -- (-3.5,0);
    \end{scope}
    \draw[dashed][line width=1.0pt](2.5,2.5) -- (2.5,3.5);
     \draw[dashed][line width=1.0pt](-1,0) -- (-1,1);
      \draw[dashed][line width=1.0pt](1.5,0) -- (1.5,1);
\end{tikzpicture}    }
\end{picture}
 \caption{Ladder sequence of diagrams for the quantity ${\cal G}$.}
 \label{fig:ladder}
 \end{figure}
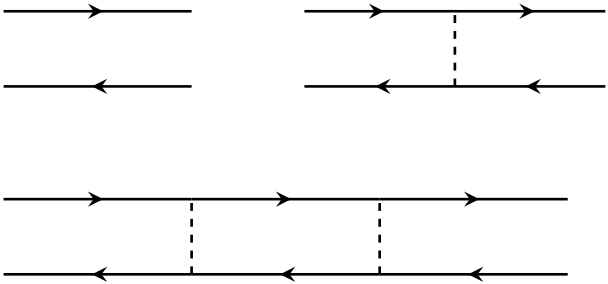

In zero approximation ${\cal G}=G(\bm r_1-\bm r_4,z)G^\star(\bm r_2-\bm r_3,z)$. Contributions to ${\cal G}$, related to the interaction term (\ref{lasem4}), can be presented by ladder diagrams, see Fig. \ref{fig:ladder}. Here a line directed to the right represents the average $G$ (\ref{prop1}) and a line directed to the left represents the average $G^\star$ (\ref{prop2}). The dotted line represents the factor $-D r^{\mu+1}$, where $r$ is the separation between the points. Summation of the ladder diagrams depicted in Fig. \ref{fig:ladder} leads to an integral equation for ${\cal G}$, analysed in Appendix \ref{sec:ladder}. Solving the integral equation, one obtains the expression (\ref{tord5}) obtained above by other method. Let us stress that the expression is the exact one in our setup due to the shortness of the refractive index correlations in $z$.

Let us analyze momenta of the intensity $I$. In the framework of our scheme, the moment $\langle I^n \rangle$ can be written as
 \begin{eqnarray}
 \langle I^n \rangle=
 \int d^2x_1 \dots d^2x_{2n}
 \left\langle \left[\Psi(0,z) \Psi^\star(0,z)\right]^n \right.
 \nonumber \\
 \left. P(\bm x_1,0) \dots P(\bm x_n,0)
 P^\star(\bm x_{n+1},0) \dots P^\star(\bm x_{2n},0) \right\rangle
 \nonumber \\
 \Psi_{in}(\bm x_1) \dots \Psi_{in}(\bm x_{2n}) .
 \label{dial1}
 \end{eqnarray}
The problem is how to calculate the average in the expression (\ref{dial1}).

Below we use the ladder approximation, where the average in the expression (\ref{dial1}) is reduced to a product of factors corresponding to the ladder diagrams. Then one finds
 \begin{eqnarray}
 \langle I^n \rangle=
 \int d^2x_1 \dots d^2x_{2n} \bigl[
 {\cal G}(0,0,\bm x_1,\bm x_{n+1},z)
 \dots
 \nonumber \\
 {\cal G}(0,0,\bm x_n,\bm x_{2n},z)
 +\dots \bigr]
 \Psi_{in}(\bm x_1) \dots \Psi_{in}(\bm x_{2n}),
 \label{powerl}
 \end{eqnarray}
where the number of summands is $n!$. Each summand produces $\langle I\rangle^n$ and we find $\langle I^n \rangle= n!\langle I\rangle^n$. In other words, we arrive at the Poisson statistics for $I$ with the probability density ${\cal P}(I)=\langle I \rangle^{-1} \exp(-I/\langle I \rangle)$. The validity of the ladder approximation should be checked separately.

Correlation functions of $I$ can be analyzed in the same ladder approximation. If we separate the points in the correlation function $\langle I(\bm r_1,z) I(\bm r_2,z) \dots \rangle$, a part of the ladders are switched off when the separation becomes larger than $R_{ph}$. The corresponding analysis is analogous to one made for the pair correlation function $\langle\Psi \Psi^\star\rangle$. Thus cumulants (irreducible parts) of the averages like $\langle I^n(\bm r_1,z) I^n(\bm r_2,z)\rangle$ becomes parametrically smaller where $\bm r=\bm r_1-\bm r_2$ exceeds $r_{ph}$.

 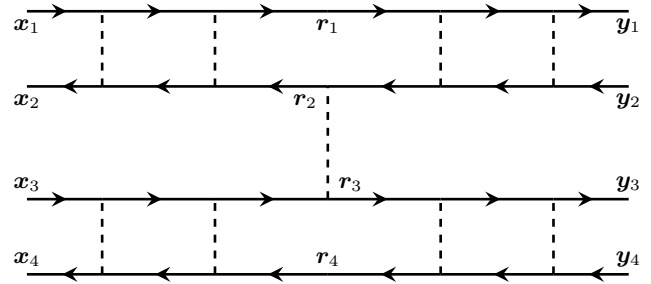
\begin{figure}
 \begin{picture}(200,110)
 \put(-20,0){ \begin{tikzpicture}
        \node at (4.5,3.3){$\bm y_1$};
        \node at (4.5,2.3){$\bm y_2$};
        \node at (4.5,1.2){$\bm y_3$};
        \node at (4.5,0.2){$\bm y_4$};
        \node at (-3.5,3.3){$\bm x_1$};
        \node at (-3.5,2.3){$\bm x_2$};
        \node at (-3.5,1.2){$\bm x_3$};
        \node at (-3.5,0.2){$\bm x_4$};
        \node at (0.5,3.3){$\bm r_1$};
        \node at (0.2,2.3){$\bm r_2$};
        \node at (0.8,1.2){$\bm r_3$};
        \node at (0.5,0.2){$\bm r_4$};
\begin{scope}[line width=1.0pt,decoration={
            markings,
            mark=at position 0.53 with {\arrow[scale=1.25,black]{stealth}}},
            ]
    \draw[postaction={decorate}] (-3.5,1) -- (-2.5,1);
    \draw[postaction={decorate}] (-2.5,1) -- (-1,1);
    \draw[postaction={decorate}](-1,1) -- (0.5,1);
    \draw[postaction={decorate}] (0.5,1) -- (2,1);
    \draw[postaction={decorate}] (2,1) -- (3.5,1);
    \draw[postaction={decorate}] (3.5,1) -- (4.5,1);
    \draw[postaction={decorate}](4.5,0) -- (3.5,0);
    \draw[postaction={decorate}] (3.5,0) -- (2,0);
    \draw[postaction={decorate}](2,0) -- (0.5,0);
    \draw[postaction={decorate}](0.5,0) -- (-1,0);
    \draw[postaction={decorate}] (-1,0) -- (-2.5,0);
    \draw[postaction={decorate}](-2.5,0) -- (-3.5,0);
\end{scope}
    \draw[dashed][line width=1.0pt](-2.5,0) -- (-2.5,1);
    \draw[dashed][line width=1.0pt](-1,0) -- (-1,1);
    \draw[dashed][line width=1.0pt](2,0) -- (2,1);
    \draw[dashed][line width=1.0pt](3.5,0) -- (3.5,1);
    \draw[dashed][line width=1.0pt](0.5,1) -- (0.5,2.5);
\begin{scope}[line width=1.0pt,decoration={
            markings,
            mark=at position 0.53 with {\arrow[scale=1.25,black]{stealth}}},
            ]
    \draw[postaction={decorate}] (-3.5,3.5) -- (-2.5,3.5);
    \draw[postaction={decorate}] (-2.5,3.5) -- (-1,3.5);
    \draw[postaction={decorate}](-1,3.5) -- (0.5,3.5);
    \draw[postaction={decorate}] (0.5,3.5) -- (2,3.5);
    \draw[postaction={decorate}] (2,3.5) -- (3.5,3.5);
    \draw[postaction={decorate}] (3.5,3.5) -- (4.5,3.5);
    \draw[postaction={decorate}](4.5,2.5) -- (3.5,2.5);
    \draw[postaction={decorate}] (3.5,2.5) -- (2,2.5);
    \draw[postaction={decorate}](2,2.5) -- (0.5,2.5);
    \draw[postaction={decorate}](0.5,2.5) -- (-1,2.5);
    \draw[postaction={decorate}] (-1,2.5) -- (-2.5,2.5);
    \draw[postaction={decorate}](-2.5,2.5) -- (-3.5,2.5);
\end{scope}
    \draw[dashed][line width=1.0pt](-2.5,2.5) -- (-2.5,3.5);
    \draw[dashed][line width=1.0pt](-1,2.5) -- (-1,3.5);
    \draw[dashed][line width=1.0pt](2,2.5) -- (2,3.5);
    \draw[dashed][line width=1.0pt](3.5,2.5) -- (3.5,3.5);
\end{tikzpicture}    }
\end{picture}
 \caption{Crossbar connecting two ladders.}
 \label{fig:crossbar}
 \end{figure}

Let us now analyze corrections to the ladder approximation. The first correction is determined by the diagrams including a crossbar connecting two ladders, see Fig. \ref{fig:crossbar}. Corrections to the averages like $\langle I^n \rangle$ are analyzed in Appendix \ref{sec:crossbar}. It is shown there that the small parameter justifying the applicability condition of the ladder approximation is $\alpha$ (\ref{smallpar}). Now we can estimate the region of applicability of the Poisson approximation. If we analyze $\langle I^n\rangle$, then the relative correction, associated with the crossbar is estimated as $n^2 \alpha$, for large $n$. Thus the Poisson expression is valid if $n \ll 1/\sqrt{\alpha}$. In other words, the exponential probability distribution is correct if $I\ll \langle I \rangle /\sqrt{\alpha}$.

Due to the refraction index fluctuations, the beam cross-section is divided into a large number of speckles $N$. The number of the speckles in the pattern can be estimated as the ratio of the beam cross-section area to the square of the correlation length (\ref{phasel}). Thus we obtain
 \begin{eqnarray}
 N\sim D^{4/(\mu+1)}z^{2+4/(\mu+1)}, \qquad
 z \gg z_\star.
 \label{speckle2}
 \end{eqnarray}
One can think about statistics of the beam intensity in the brightest speckle. For this purpose, we assume that the intensity statistics in the speckles are independent and are determined by the Poisson statistics. The corresponding analysis is made in Appendix \ref{sec:maximum}. The main result of the analysis is that the average intensity in the brightest speckle can be estimated as $\langle I \rangle \ln N$. Due to slowness of the logarithmic function this value remains inside the applicability region of Poisson statistics determined by the condition $I\ll \langle I \rangle /\sqrt{\alpha}$. That justifies our conclusions.

\section{Conclusion}
\label{sec:Conclusion}

We analyzed statistical properties of the speckle pattern of light intensity  in the cross-section of the laser beam  propagating in the turbulent fluid (atmosphere). The pattern is formed due to refraction index fluctuations caused by the turbulent fluctuations. We demonstrated that there are two characteristic dimensionless propagation length, $z_{rytov}$ and $z_\star$, determined by the relations (\ref{rytovdimensionless}) and (\ref{star}). Our theory is valid under the condition $z_{rytov}\lesssim z_\star$. We are interested in the region $z\gg z_\star$ where effects of the random diffraction dominate.

If the propagation length $z$ satisfies the condition $z \gg z_\star,$ then the beam contains many speckles and its width $r_{width}$ becomes much wider than in the case of the free diffraction propagation. This is due to the refractive index fluctuations. The beam width is determined by the expression (\ref{beamwidth}), that is
 \begin{equation*}
 r_{width}\propto z^{1/(\mu+1)+1}.
 \end{equation*}
The correlation length of the signal $r_{ph}$, related to fluctuations of its phase, appears to be much smaller than  the beam width and can be estimated in accordance with Eq. (\ref{phasel}) by
 \begin{equation*}
 r_{ph}\propto z^{-1/(\mu+1)}.
 \end{equation*}

We found the analytical expression for the pair correlation function (\ref{pairc4}) in the region $z \gg z_\star$ and demonstrated that higher order correlation functions of the envelope $\Psi$ are split into products of the pair correlation functions. This result is in accordance with the
expectation that strong phase fluctuations lead to an effective Gaussianity of the envelope statistics. We analyzed also non-Gaussian corrections to the higher order correlation functions and established that they are controlled by the $z$-dependent parameter $\alpha\propto z^{\mu+1-4/(\mu+1)}$, see Eq. (\ref{smallpar}). As one expects, the parameter diminishes as $z$ grows due to the increasing role of the random diffraction.

We developed also the diagrammatic technique for calculating corrections
to the correlations functions of the envelope related to the random
diffraction. In the diagrammatic language, the effective Gaussianity of
the envelope statistics is explained as the approximation where the
so-called ladder diagrams are taken into account. Let us stress that the
approximation implies a deep resummation of the diagrams. The diagrammatic
technique gives a powerful tool to go beyond the scope of the Gaussian
approximation and enables one to calculate analytically non-Gaussian
corrections.

The Gaussianity of the envelope statistics results in the Poisson statistics of
the beam intensity $I$. We established the applicability region of the
statistics  is given by  $I\ll \langle I \rangle /\sqrt{\alpha}$, where $\langle
I \rangle$ is the average intensity inside the pattern and $\alpha$ is the
parameter (\ref{smallpar}). We examined the statistics of the brightest
spot among the large number $N\sim (r_{width}/r_{ph})^2$ in the speckle
pattern. The average value of the intensity inside the brightest spot can
be estimated as $\langle I \rangle \ln N$. The quantity lies inside the
applicability region of the Poisson approximation.

In the opposite limit $z_{rytov}\gg z_\star$, compare to considered in this paper, there appears an intermediate region $z_{rytov}\gg z\gg z_\star$, of the dimensionless propagation lengths $z$, where fluctuation effects are extremely strong. The approach developed in this paper is not applicable to that region, being, however, applicable to the region $z\gg z_{rytov}$. The region $z_{rytov}\gg z\gg z_\star$ needs a special analysis based on the diagrammatic technique and the related quantum field
theory. Such analysis is outside of the scope of this paper.

\begin{acknowledgements}

We thank support of the Russian Ministry of Science and High Education,
program 0033-2019-0003. The  work of P.M.L. was   supported by the
state assignment ``Dynamics of the complex systems". The work of
P.M.L.  was   supported by the National Science Foundation, grant
DMS-1814619. Simulations were performed  at the Texas Advanced
Computing Center using the Extreme Science and Engineering Discovery
Environment (XSEDE), supported by NSF Grant ACI-1053575.

\end{acknowledgements}

\appendix

\section{Ladder representation}
\label{sec:ladder}

Here, we demonstrate how to obtain the expression (\ref{tord5}) for the Green function (\ref{dia1}) using the ladder representation for the object. The quantity ${\cal G}$ is depicted by the sum of the ladder sequence of the diagrams depicted in Fig. \ref{fig:ladder}. The ladder representation leads to the following integral equation
 \begin{eqnarray}
 {\cal G}(\bm r_1, \bm r_2, \bm r_3, \bm r_4,z)
 =G(\bm r_1-\bm r_4,z) G^\star(\bm r_2-\bm r_3,z)
 \nonumber \\
 -\int d\zeta \, d^2 r_5 d^2 r_6 D |\bm r_{56}|^{\mu+1}
 G(\bm r_{15},z-\zeta) G^\star(\bm r_{26},z-\zeta)
 \nonumber \\
 {\cal G}(\bm r_5, \bm r_6, \bm r_3, \bm r_4,\zeta). \qquad
 \label{ladf1}
 \end{eqnarray}
Using the relation
 \begin{eqnarray}
 (\partial_z -i\nabla_1^2+i\nabla_2^2)
 [G(\bm r_1,z)G^\star(\bm r_2,z)]
 \nonumber \\
 =\delta(z)\delta(\bm r_1)\delta(\bm r_2).
 \nonumber
 \end{eqnarray}
following from Eqs. (\ref{prop1},\ref{prop2}), one obtains from Eq. (\ref{ladf1})
 \begin{eqnarray}
 (\partial_z -i\nabla_1^2+i\nabla_2^2+D r_{12}^{\mu+1}){\cal G}
 \nonumber \\
 =\delta(z)\delta(\bm r_1-\bm r_4)\delta(\bm r_2-\bm r_3).
 \label{ladf2}
 \end{eqnarray}
Passing to the variables $\bm R=(\bm r_1+\bm r_2)/2$, $\bm r=\bm r_1-\bm r_2$, we rewrite the equation (\ref{ladf2}) as
 \begin{eqnarray}
 \left(\partial_z -2i\frac{\partial^2}{\partial \bm r \partial \bm R}+D r^{\mu+1}\right){\cal G}
 \nonumber \\
 =\delta(z)\delta(\bm r-\bm x)\delta(\bm R-\bm X),
 \label{ladf3}
 \end{eqnarray}
where $\bm X=(\bm r_3+\bm r_4)/2$, $\bm x=\bm r_4-\bm r_3$. The equation (\ref{ladf3}) is equivalent to Eq. (\ref{tord4}).

\section{Corrections to the ladder approximation}
\label{sec:crossbar}

Here we demonstrate how to find corrections to the ladder approximation. For this purpose we consider the first correction to the product of two ladders giving the main contribution to the fourth-order correlation function of $\Psi$. The correction is determined by the diagrams of the type depicted in Fig. \ref{fig:crossbar}, containing the only crossbar connecting two ladders. The sum of the ladder diagrams of the type presented in Fig. \ref{fig:crossbar} gives the first correction to the product ${\cal G}(\bm y_1,\bm y_2,\bm x_2,\bm x_1,z){\cal G}(\bm y_3,\bm y_4,\bm x_4,\bm x_3,z)$.

After summation of the ladder sequences, the diagrams depicted in Fig. \ref{fig:crossbar} give the following analytical expression
 \begin{eqnarray}
 -D\int d\zeta\ d^2r_1 d^2r_2 d^2 r_3 d^2r_4 |\bm r_2-\bm r_3|^{\mu+1}
 \nonumber \\
 {\cal G}(\bm y_1,\bm y_2,\bm r_2,\bm r_1,z-\zeta)
 {\cal G}(\bm y_3,\bm y_4,\bm r_4,\bm r_3,z-\zeta)
 \nonumber \\
 {\cal G}(\bm r_1,\bm r_2,\bm x_2,\bm x_1,\zeta)
 {\cal G}(\bm r_3,\bm r_4,\bm x_4,\bm x_3,\zeta),
 \nonumber
 \end{eqnarray}
where the integral over $\zeta$ goes from $0$ to $z$. In derivation of the expression we used the relations
 \begin{eqnarray}
 \int d^2x\, G(\zeta,\bm x) G(z-\zeta, \bm r-\bm x)
 =-i G(z,\bm r),
 \nonumber \\
 \int d^2x\, G^\star(\zeta,\bm x) G^\star(z-\zeta, \bm r-\bm x)
 =i G^\star(z,\bm r),
 \label{prop10}
 \end{eqnarray}
for the functions (\ref{prop1},\ref{prop2}), that can be checked directly.

Taking into account the expression (\ref{lasem4}) for the interaction and the relations (\ref{prop10}), one finds ultimately the following correction $\Theta$ to the product ${\cal G}(\bm y_1,\bm y_2,\bm x_2,\bm x_1,z){\cal G}(\bm y_3,\bm y_4,\bm x_4,\bm x_3,z)$
 \begin{eqnarray}
 \Theta=-D\int d\zeta\ d^2r_1 d^2r_2 d^2 r_3 d^2r_4
 \left(|\bm r_2-\bm r_3|^{\mu+1} \right.
 \nonumber \\  \left.
 +|\bm r_1-\bm r_4|^{\mu+1}
 -|\bm r_1-\bm r_3|^{\mu+1}-|\bm r_2-\bm r_4|^{\mu+1}\right)
 \nonumber \\
 {\cal G}(\bm y_1,\bm y_2,\bm r_2,\bm r_1,z-\zeta)
 {\cal G}(\bm y_3,\bm y_4,\bm r_4,\bm r_3,z-\zeta)
 \nonumber \\
 {\cal G}(\bm r_1,\bm r_2,\bm x_2,\bm x_1,\zeta)
 {\cal G}(\bm r_3,\bm r_4,\bm x_4,\bm x_3,\zeta),
 \label{prop11}
 \end{eqnarray}

Let us consider the case $\bm x_i=\bm y_i=0$. We introduce the quantities
 \begin{eqnarray}
 \bm r_1=\bm X +\bm R/2 +\bm y/2, \
 \bm r_2=\bm X +\bm R/2 -\bm y/2,
 \nonumber \\
 \bm r_3=\bm X -\bm R/2 +\bm s/2, \
 \bm r_4=\bm X -\bm R/2 -\bm s/2,
 \label{varia}
 \end{eqnarray}
where $\bm s$ and $\bm y$ measure the ladder thickness whereas $\bm R$ is the separation between the ladders. Taking the expression (\ref{tord5}) into account, we conclude that integration over $\bm X$ in Eq. (\ref{prop11}) produces a $\delta$-function fixing $\bm y=-\bm s$. After integrating over $\bm s$, one arrives at the expression
 \begin{eqnarray}
 \Theta=\frac{D}{2^{14}\pi^6}\int \frac{d\zeta\ d^2R\, d^2y}{z^2 \zeta^2 (z-\zeta)^2}
 \nonumber \\
 \left(|\bm R+\bm y|^{\mu+1} +|\bm R-\bm y|^{\mu+1}-2R^{\mu+1}\right)
 \nonumber \\
 \exp\left[ \frac{iz}{2\zeta (z-\zeta)}\bm y \bm R
 -\frac{2Dz}{2+\mu}y^{\mu+1}\right].
 \label{prop12}
 \end{eqnarray}

One can take the integral over $\bm R$ in the expression (\ref{prop12}) explicitly to obtain
that \begin{eqnarray}
 \Theta=\frac{2^{2c} c D}{2^{8} \pi^4 z^{3-c}}
 \frac{\Gamma(2+\mu/2)}{\Gamma(1-c/2)}
 \int_0^1d\chi\, [\chi(1-\chi)]^{\mu+1}
 \nonumber \\
 \int_0^\infty \frac{d y}{y^{2+\mu}}
 \exp\left(-\frac{2Dzy^{\mu+1}}{2+\mu}\right)
 \sin^2\left[\frac{y^2}{4z\chi(1-\chi)}\right].
 \label{prop14}
 \end{eqnarray}
One can worry about a singular contribution related to small $\chi(1-\chi)$. At $y^2\lesssim z\chi(1-\chi)$ the integral (\ref{prop14}) converges. Thus, at small $\chi(1-\chi)$ the integral over $y$ produces a singular contribution $\propto [\chi(1-\chi)]^{-(\mu+1)/2}$, suppressed by the factor $[\chi(1-\chi)]^{\mu+1}$ in Eq. (\ref{prop14}). Thus we arrive at the natural estimates $\zeta\sim z$, $y\sim (Dz)^{-1/{\mu+1}}$, $R\sim D^{1/(\mu+1)}z^{1+1/(\mu+1)}\gg y$. Therefore the integral (\ref{prop14}) is estimated as
 \begin{equation}
 \Theta\sim \frac{D}{z^{5-c}} y^{3-\mu}
 \sim \frac{1}{z^4}\alpha.
 \label{prop15}
 \end{equation}
The factor $\alpha$ (\ref{smallpar}) characterizes smallness of the corrections to the ladder diagrams.

Let us examine the pair correlation function $\langle I I \rangle$ at distances much larger than the correlation length  $R_{ph}$ (\ref{phasel}). The main contribution to the irreducible part of the correlation function is determined by the same diagram depicted in Fig. \ref{fig:crossbar}.  If $z\ll z_\star$, then $\bm x_i$ are of order of $a$ and can be neglected. Then $\bm y=-\bm s\sim (Dz)^{-1/(\mu+1)}$ and we arrive at the same smallness (\ref{smallpar}). If $z\gg z_\star$ then $x_{12}\sim x_{34}\sim (Dz)^{-1/(\mu+1)}$ and $y\sim s\sim (Dz)^{-1/(\mu+1)}$ as well. And we return to the same smallness (\ref{smallpar}). The correlation length of the correction to the pair correlation function $\langle I I \rangle$ is $r_{width}$ (\ref{tord8}). An analogous analysis can be done for the other correlation functions of $I$.

\section{Statistics of a largest value}
\label{sec:maximum}

We examine here the case where independent variables $x_1,\dots, x_{N+1}$ have identical Poisson probability density functions
 \begin{equation}
\label{pu1}
{\cal P}(x_j)=\frac{1}{q}\exp\left(-\frac{x_j}{q}\right), \quad \langle x_j \rangle=q.
\end{equation}
We are interested in the statistics of the largest value in the sequence $x_1,\dots, x_{N+1}$. Let us designate the largest value as $y$. Then the probability density function of $y$ can be written as
 \begin{eqnarray}
 {\cal P}(y)=\frac{N+1}{q}\exp\left(-\frac{y}{q}\right)
 \left[\frac{1}{q}\int\limits_{0}^{y}dx\,e^{-x/q}\right]^N
 \label{su1} \\
 =\frac{N+1}{q}\exp\left(-\frac{y}{q}\right) \left(1-e^{-y/q}\right)^N.
 \nonumber
 \end{eqnarray}
Further we assume $N\gg 1$ and, consequently, the average value $\langle y\rangle \gg q$. In this case it is possible to rewrite the expression (\ref{su1}) as
\begin{eqnarray}
{\cal P}(y)\approx \frac{N}{q}
\exp\left(-\frac{y}{q}-Ne^{-y/q}\right).
 \label{su2}
\end{eqnarray}
Then it is possible to calculate easily the average value
 \begin{eqnarray}
 \langle y\rangle =\int dy\ y{\cal P}(y)
 \approx q\ln N.
 \label{su3}
\end{eqnarray}
We see that the average value $\langle y\rangle$ (\ref{su3}) is much larger than $q$, indeed, since $N\gg 1$. That justifies our conclusions.

\section{Bright speckles}

Here we consider bright speckles, that are characterized by the inequality $I\gg \langle |\Psi(\bm y)|^2 \rangle$. Here $\bm y$ is the center of the speckle and $I$ is the beam intensity at the point. We are interested in the shape of such bright speckle. To solve the problem, one can use the saddle-point approach. However, we assume that we are still inside the Gaussian approximation.

Let us analyze the probability density $P$, that at some point $\bm y$ the beam intensity is $I$. In the Gaussian approximation the probability density can be written as
 \begin{eqnarray}
 P(I,\bm y)=\int D\Psi\ D\Psi^\star\ {\cal N}
 e^{-{\cal S}}
 \delta\left[\Psi^\star(\bm y) \Psi(\bm y)-I\right],
 \label{bright1}
 \end{eqnarray}
where ${\cal N}$ is the normalization factor and ${\cal S}$ is the effective action. It is written as
 \begin{equation}
 {\cal S}=\int d^2 r\ \Psi^\star \hat{K} \Psi,
 \label{bright6}
 \end{equation}
where $\hat{K}$ is the operator, related to the pair correlation function:
 \begin{equation}
 \hat K F_2(\bm r_1,\bm r_2)=\delta(\bm r_1-\bm r_2).
 \label{bright2}
 \end{equation}
Obviously, $\int dI\ P(I,\bm y)=1$.

We rewrite the expression (\ref{bright1}) as
 \begin{eqnarray}
 P(I,\bm y)=\int D\Psi\ D\Psi^\star\ {\cal N} \int \frac{d\lambda}{2\pi i}
 \nonumber \\
 \exp\left[-\int d^2 r\ \Psi^\star \hat{K} \Psi
 +\lambda \Psi^\star(\bm y) \Psi (\bm y) -\lambda I \right],
 \label{bright3}
 \end{eqnarray}
where integration over $\lambda$ goes along the imaginary axis. If $I$ is high, then the integration in the expression (\ref{bright3}) can be performed in the saddle-point approximation. The saddle-point equation for $\Psi$ is
 \begin{equation}
 \hat K \Psi=\lambda \Psi(\bm y) \delta(\bm r-\bm y).
 \label{bright4}
 \end{equation}
Taking into account the equation (\ref{bright2}), one finds from Eq. (\ref{bright4})
 \begin{equation}
 \Psi(\bm r)=\lambda \Psi(\bm y) F_2(\bm r,\bm y).
 \label{bright5}
 \end{equation}
Therefore
 \begin{equation}
 \lambda F_2(\bm y,\bm y)=1.
 \label{bright7}
 \end{equation}
Multiplying the relation (\ref{bright5}) by $\Psi^\star(\bm y)$, one obtains
 \begin{equation}
 \Psi(\bm r)\Psi^\star(\bm y)=\lambda I F_2(\bm r,\bm y).
 \label{bright8}
 \end{equation}
where the self-consistency condition $\Psi^\star(\bm y) \Psi(\bm y)=I$ is used. As we see from Eq. (\ref{bright8}), the saddle-point profile is determined by the pair correlation function.

As it follows from Eqs. (\ref{bright3}) and (\ref{bright4}), the saddle-point value of the probability density $P(I,\bm y)$ is $\exp(-\lambda I)$. Using the relation (\ref{bright7}), one obtains
 \begin{equation}
 P(I,\bm y)\sim \exp\left[-\frac{I}{F_2(\bm y,\bm y)}\right]
 =\exp\left[-\frac{I}{\langle |\Psi(\bm y)|^2 \rangle}\right].
 \label{bright9}
 \end{equation}
Thus we return to the Poisson statistics. In addition, we see that the applicability condition of the saddle-point approximation is $I\gg \langle |\Psi(\bm y)|^2 \rangle$, indeed.

\end{document}